\documentclass[aps,prd,twocolumn,showpacs,groupedaddress,longbibliography,preprintnumbers]{revtex4-1}

\pdfoutput=1
\usepackage{graphicx} 
\usepackage{amsmath}  
  \newcounter{runningcount}
\usepackage{hyperref}
\usepackage{breakurl}

\newcommand{\tev}{\hbox{ TeV}}
\newcommand{\gev}{\hbox{ GeV}}
\newcommand{\fb}{\hbox{ fb}}
\newcommand{\lum}{\hbox{ cm}^{-2}\hbox{s}^{-1}}
\newcommand{\alphas}{\ensuremath{\alpha_{\mathrm{s}}}}
\newcommand{\orcid}[1]{\thanks{\href{http://orcid.org/#1}{ORCID: #1}}}

\begin{document}
\title{Dream Machines}%
\preprint{FERMILAB-PUB-18-305-T}
\author{Chris Quigg}
\email[]{quigg@fnal.gov}
\orcid{0000-0002-2728-2445}
\affiliation{Fermi National Accelerator Laboratory \\ P.O.\ Box 500, Batavia, Illinois 60510 USA}
\date{8 November 2018} 

\vspace*{6pt}
\begin{abstract}
\centerline{\fbox{\emph{Dedicated to the memory of \hbox{Burton Richter}---master of the craft   and exemplary citizen-scientist}}}

\vspace*{6pt}
\noindent
Particle accelerators and their detectors are the world's most powerful microscopes. They enable us to inspect the constituents of matter at attometer scales, study matter under unusual conditions, and concentrate extraordinary amounts of energy into tiny volumes to create new forms of matter and initiate new phenomena. The progress of particle physics and of accelerator science and technology go hand in hand. I look to the past, present, and future, raising questions that we would like to answer about nature along the way. \end{abstract}
\maketitle


\section{Invitation}

\vspace*{-12pt}
\noindent{The romance of the great accelerators} helped attract me to particle physics, and my sense of wonder that human beings can create these remarkable devices and use them to interrogate nature has only grown over time. As an undergraduate student, I read that scientists on the West Coast were going to collide electron beams head-on, and they didn't know what would happen. How wonderful!---to not know the answer in advance, but to devise instruments that make it possible to find out~\cite{Barber:1966kd}. (To a young boy who knew next to nothing, it was much more intriguing to hear that ``they don't know what will happen'' than that they are going to test Quantum Electrodynamics.) 
 This brief essay solicits answers that I do not know, along with some musings about future machines that might lead us to new understanding.

In the Preface to his brief history of accelerators, published in 1969, Stanley Livingston wrote~\cite{Livingston:PABH},
\begin{quote}
``Particle accelerators are among the most useful tools for research in nuclear physics and in high-energy particle physics. The rapid growth of these research fields has been due, in large measure, to the development of a sequence of electronuclear machines for acceleration of ions and electrons. The high-intensity and well-controlled beams from these machines can be used to disintegrate nuclei, produce new unstable isotopes, and investigate the properties of the nuclear force. Modern high-energy accelerators can produce excited states of elementary particles of matter, forming new unstable particles with mass values much higher than those of the stable particles. Fundamental questions can be asked of nature, and answered by experiments with these very high-energy particles. The field of high-energy particle physics is on the threshold of a significant breakthrough in our understanding of the particles of nature and the origins of the nuclear force.''
\end{quote}
Livingston's assessment has continuing relevance. Nuclear physics and particle physics have become increasingly intertwined thanks to the rise of the standard model as a universal language, and astroparticle physics has emerged as a companion to both. Accelerator-based experiments have made decisive contributions, as high-luminosity colliders have joined fixed-target machines. Livingston's generic ``ions and electrons'' now include electrons and positrons, protons and antiprotons, a variety of relativistic heavy ions, and may some day encompass muons. Superconducting technology for magnets and radio-frequency accelerating cavities, active optics and beam-cooling techniques that build on the great innovation of strong focusing, large-scale cryogenics and vacuum technology---these are among the advances that have enabled a versatile spectrum of machines. 

Much of the history of our field over recent decades has been written in experiments at Brookhaven's Alternating Gradient Synchrotron, the Budker Institute in Novosibirsk, CERN's Proton Synchrotron, Intersecting Storage Rings, and Super Proton Synchrotron, the U-70 proton synchrotron in Protvino, Fermilab's Tevatron and Main Injector, HERA at DESY, the Japan Proton Accelerator Research Complex, the Stanford Linear Collider, LEP, flavor factories at Beijing, Cornell, DESY, Frascati, KEK, and SLAC, and the Large Hadron Collider at CERN~\cite{[{The characteristics of many accelerators are presented in }][]{Handbook}}. Since Livingston's day, we have progressed from photographic emulsions, bubble chambers, and spark chambers to the highly evolved detectors that incorporate many advanced technologies at different scales. Peter Galison gives an overview in his historical survey, \emph{Image and Logic}~\cite{ImageLogic}. The LHC detectors~\cite{Froidevaux2006} and upgrades planned for the (high-luminosity) HL-LHC~\cite{Campana2016} represent the greatest complexity and performance yet achieved, while the developments in computing needed to keep pace are detailed in Refs.~\cite{Bird:1695401,Bauerdick:2018qjx}.
%

A revealing way to assess the progress in particle physics over a half century is to consider the Problems of High-Energy Physics cited in the (Fermi) National Accelerator Laboratory Design Report~\cite{Cole:1968wx}:
\begin{quotation}{\ttfamily
Which, if any, of the particles that have so far been discovered, is, in fact, elementary, and is there any validity in the concept of "elementary" particles? 

What new particles can be made at energies that have not yet been reached? Is there some set of building blocks that is still more fundamental than the neutron and the proton? 

Is there a law that correctly predicts the existence and nature of all the particles, and if so, what is that law? 

Will the characteristics of some of the very short-lived particles appear to be different when they are produced at such higher velocities that they no longer spend their entire lives within the strong influence of the particle from which they are produced? 

Do new symmetries appear or old ones disappear for high momentum-transfer events? 

What is the connection, if any, of electromagnetism and strong interactions? 

Do the laws of electromagnetic radiation, which are now known to hold over an enormous range of lengths and frequencies, continue to hold in the wavelength domain characteristic of the subnuclear particles? 

What is the connection between the weak interaction that is associated with the massless neutrino and the strong one that acts between neutron and proton? 

Is there some new particle underlying the action of the "weak" forces, just as, in the case of the nuclear force, there are mesons, and, in the case of the electromagnetic force, there are photons? If there is not, why not? 

In more technical terms: Is local field theory valid? A failure in locality may imply a failure in our concept of space. What are the fields relevant to a correct local field theory? What are the form factors of the particles? What exactly is the explanation of the electromagnetic mass difference? Do "weak" interactions become strong at sufficiently small distances? Is the Pomeranchuk theorem true? Do the total cross sections become constant at high energy? Will new symmetries appear, or old ones disappear, at higher energy? 

}
\end{quotation}
Young physicists in particular may be a bit astonished by how little our colleagues knew in 1968 of what we consider textbook material. But consider how much more they knew than their forebears of fifty years before, and be prepared to be chastened by how quaint and incomplete our current knowledge will seem fifty years hence! While some of the questions are hard to decrypt---what, specifically, was in their heads?---overall they exhibit a great deal of insight into the kind of issues that might matter. Indeed, several   could serve in our time.
\section{So many ideas!}
Four decades ago, Robert R. Wilson, Fermilab's first Director, published ``Fantasies of future Fermilab facilities,'' an expansive essay on possible future projects for the laboratory~\cite{Wilson:1979zg}. Blending kid-in-a-candy-store enthusiasm with a measure of realism, Wilson's fantasies remind us of the value of raising possibilities, even if they cannot all be carried out as imagined. 

The most immediately achievable project was the Tevatron, the superconducting ring that was realized in the Main Ring tunnel, one kilometer in radius. In its first incarnation, protons were to be accelerated to approximately $1\tev$ for use in fixed-target experiments either directly or through the production of secondary (pions, kaons, etc.) and tertiary (electrons, photons, neutrinos) beams. Next for consideration was 250-GeV protons from the Main Ring colliding with 1-TeV protons in the Tevatron. This two-ring scenario would be followed by 1-TeV antiprotons colliding with 1-TeV protons in the Tevatron. Another variation would have been to accelerate electrons to $12\gev$ in the Main Ring, and collide those with 1-TeV protons from the Tevatron. Not all of these protoprojects materialized. Fermilab constructed the ``Super Ring'' Tevatron and operated at $800\gev$ in fixed-target mode. The proton--antiproton collider, which built on the experience of the S$p\bar{p}$S Collider at CERN, eventually attained $0.98\tev$ per beam. These served as essential instruments for particle physics over a quarter century. On the other hand, the Main Ring on Tevatron scheme was judged inadequate to the search for the electroweak gauge bosons, and a far more ambitious electron--proton collider ($27.5\gev$ on $920\gev$) was realized at DESY in Hamburg.

Wilson next took up what he called an Accumulator Ring, to be sited either in the Main Ring tunnel or in the (8-GeV) Booster tunnel, to gain a whole series of possible efficiencies. None of these took shape, but we can see in his musings the germ of the Recycler ring, based on permanent magnets, in the Main Injector tunnel. He also sketched Bypasses to enhance the compatibility of the fixed-target and collider programs; these did not come to pass, but perhaps they set the stage for the overpasses at the CDF and D0 detectors that made the Main Ring a nonplanar machine. A concentric Inner Ring 80\% of the Main Ring radius, fitted with 8-tesla magnets could contain 1.5-TeV protons or antiprotons, enabling $pp$ collisions with the Tevatron or stand-alone $\bar{p}p$ collisions.

POPAE, for Protons On Protons And Electrons, referred to a high-luminosity $1\otimes 1$-TeV proton--proton
collider with a $0.2\otimes 1$-TeV electron--proton option, a joint initiative of Argonne and Fermilab, to be housed in two rings in a 5.5-km--radius tunnel separate from the Main Ring. It did not gain community support.

Finally, it was time for Wilson to think big, with a site-filling machine that he called the Pentevac. It was to provide 5-TeV protons on fixed targets, 5-TeV antiprotons on 5-TeV protons, 50-GeV electrons on 50-GeV positrons, and 10-to-50-GeV electrons on 5-TeV protons. And it might be staged, as today we imagine a sequential approach to the FCC components at CERN. 
It is noteworthy that he imagined replacing the NbTi superconductor of the Tevatron magnets with Nb$_3$Sn---a preoccupation of magnet development today.

All of these fantasies were enunciated by a single lab director at one moment. Similarly rich thinking took place at many institutions around the world. I would draw a couple of lessons. First, not every idea---not even every good idea---becomes reality, but those that do can have decisive influence on the development of our field. Second, ideas do not die. We have a collective memory; old ideas generate new ones, or find application in new settings. I see this regularly in theoretical physics, where ideas once set aside as not especially relevant, or having run their course, reappear to very significant effect. Surely it is so for accelerator building as well.

Let's move to the present. While not typically expressed with Wilson's antic enthusiasm, many dazzling possibilities lie in front of us. The SuperKEKB project is in commissioning, intensity improvement projects motivated by long-baseline neutrino experiments are moving forward at J-PARC and Fermilab, and the Facility for Antiproton and Ion Research is under construction by the European nuclear physics community at Darmstadt. The next approved project is the HL-LHC, a very ambitious luminosity upgrade at CERN that promises event samples of $3\,000\fb^{-1}$ in $pp$ collisions at energies approaching $\sqrt{s} = 14\tev$. Beyond that, if we confine our attention to major studies that are reasonably mature, we see 
\begin{itemize}
\item The International Linear Collider, with a first phase envisioned as $e^+e^-$ collisions at $\sqrt{s}=250\gev$.
\item HE-LHC, an energy doubler for the LEP/LHC tunnel, reaching $\sqrt{s} \approx 27\tev$ for $pp$ collisions.
\item CLIC-380, a first phase of the Compact LInear Collider for CERN, with $e^+e^-$ collisions up to $\sqrt{s}=380\gev$.
\item LH$e$C, to collide a 60-GeV electron beam with the LHC proton beam.
\item An Electron--Ion Collider, developed by the nuclear physics community in the United States. The Brookhaven design calls for 30-GeV electrons in collision with 100-GeV/u ions or 250-GeV protons. Jefferson Lab has put forward a staged approach, beginning with 12-GeV electrons on 20-GeV/u ions or 40-GeV protons, and leading to 20-GeV electrons on 40-GeV/u ions or 100-GeV protons.
\item Three colliders under the Future Circular Colliders initiative based at CERN, which explores the possibilities for a tunnel approximately 100 kilometers in circumference. The baseline parameters of energy and luminosity for the hadron collider, FCC-hh, are $\sqrt{s} = 100\tev$, $\mathcal{L} \to 3 \times 10^{35}\lum$ for $pp$ collisions. The ``Higgs-factory'' element, FCC-ee, could operate at high (energy-dependent) luminosity from $\sqrt{s} \approx M_Z$ to $365\gev$, past top-pair threshold. The electron-hadron element, FCC-eh, projects concurrent operation with FCC-hh, colliding 60-GeV electrons with the proton or ion beam.
\item A  group centered at IHEP (Beijing) is developing a Conceptual Design Report for the Circular Electron--Positron Collider (CEPC) and its proposed successor, the Super Proton-Proton Collider (SppC). Like the FCC study, CEPC--SppC is based on a very large ring that could accommodate frontier machines in sequence.
\end{itemize}
All of these have considerable scientific merit. That is why they have attracted numbers of scientists to make serious studies of accelerator, detector, and scientific promise. They are all, in varying degrees, large, costly, challenging, and distant in time.
Beyond this list, we see studies of muon storage rings and muon colliders, a so-called photon collider,  multi-TeV lepton colliders, high-energy $\beta$ beams as specialized neutrino sources, and concepts for particle-physics experiments at spallation sources.

\emph{We will not execute all of these ideas.} They will compete for resources, and for our enthusiasm. But it is  far better to have too many appealing ideas than too few! Let us now look at a few of the scientific imperatives that motivate future accelerators, and that will help us formulate a  Planetary Program for Particle Physics.

\section{Beyond the Higgs-boson discovery}
The discovery of the Higgs boson by the ATLAS and CMS Collaborations working at the LHC is a landmark for our understanding of nature and a remarkable achievement of many people---especially the experimenters and accelerator builders whose extended effort made the discovery happen~\cite{[{My before-and-after papers, }][{, provide some intellectual history, review of theoretical contributions, and context.}]Quigg:2009vq}. We can say of the new, unstable particle with mass $M_H = 125.18 \pm 0.16\gev$ that the evidence is developing as it would for a textbook Higgs boson of the standard electroweak theory. Its observed decays into $W^+W^-$ and $ZZ$ implicate $H(125)$ as an agent of electroweak symmetry breaking. It decays to $\gamma\gamma$ at approximately the expected rate. It is dominantly spin-parity $J^P=0^+$. Evidence for the $Ht\bar{t}$ coupling from the dominant production mechanism, gluon fusion through a top-quark loop, and from the recent observation of $t\bar{t}H$ production, imply that the Higgs field plays a role in the generation of fermion masses. That implication is supported by the observation of decays into $\tau^+\tau^-$ and $b\bar{b}$ at roughly the expected rates. We have seen that the $\mu^+\mu^-$ decay rate is suppressed (the dimuon channel hasn't yet been observed), as would be expected if the $H$ couplings to quarks and leptons scale with fermion mass. The LHC experiments are sensitive to the gluon-fusion and vector-boson production mechanisms, associated production of $H$ + an electroweak gauge boson, and $Ht\bar{t}$ production. At the current precision, the observed yields, which measure production cross section times branching ratios, are in line with standard-model expectations. We have found no evidence yet for charged or neutral companions to $H(125)$, and no suggestion of new strong dynamics. Although there is no direct measurement of the Higgs-boson width, deft analysis of interference effects very plausibly bounds the width at less than about four times the standard-model expectation.

This already impressive dossier indicates that the prospecting (or search-and-discovery) phase is over, and that we are moving to a painstaking forensic investigation. What remains to be learned? So far, we have only seen couplings between $H$ and ``third-generation'' fermions, $t, b, \tau$. It is important to test whether the Higgs field is also the giver of mass to the lighter quarks and to the charged leptons $\mu$ and $e$. If $H \to \mu^+\mu^-$ occurs at the standard-model rate, it will surely be observed in the $3\,000\fb^{-1}$ sample of the HL-LHC, and might well be established in current running aimed at $300\fb^{-1}$. This is an important branch point for theories of the fermion masses: it would rule against pictures in which the source of light-fermion masses is quantum effects tied to the heavy fermions. Verifying the standard-model $H\mu\bar{\mu}$ coupling would raise interest in a $\mu^+\mu^- \to H$ factory. Some studies indicate that a muon collider's beam-momentum spread might be fine enough to permit mapping out the $H$ line shape, which would be a most impressive feat. Establishing the origin of the electron's mass occupies a special place in our quest to understand the nature of matter. The electron mass sets the scale of the Bohr radius, and so the size and integrity of atoms, prerequisites to valence bonding. If I ruled the Universe, I would award the Nobel Prize in Chemistry to whoever figures this out. Showing that the Higgs field is responsible is no easy task: the $H\to e^+e^-$ branching fraction  is only about five parts per billion. FCC-ee enthusiasts express hope that the formation reaction $e^+e^- \to H$ might be in reach, but we are far from knowing that it could be done. 

The LHC experiments access several production channels and many decay modes; they will provide much additional precise information. But it would be advantageous to have a second look through the reaction $e^+e^- \to HZ$. If a high-luminosity $e^+e^-$ Higgs factory were to drop out of the sky tomorrow, the line of users would be very long. Although that marvelous event will not happen, several ambitious proposals are in view. Their particular assets include the ability to determine absolute branching fractions and to measure directly the total width of the Higgs boson, and graceful access to decay channels such as $H \to c\bar{c}$. Important complementary information could be gleaned in other operating modes: Tera-$Z$ for both precision tests and discovery, a $WW$ threshold scan, and studies at $t\bar{t}$ threshold. A more comprehensive set of precise measurements, from the LHC, HL-LHC, and future machines, will enable us to make ever more incisive tests of the standard model as a quantum field theory, and to reflect on the implications of $M_H \approx 125\gev$.

I close this section with a list of questions about electroweak symmetry breaking and the Higgs sector that we must answer to approach a final verdict about how closely $H(125)$ matches the textbook Higgs boson.
\begin{enumerate}
\item Is $H(125)$ the only member of its clan? Might there be others---charged or neutral---at higher or lower masses?
\item Does $H(125)$ fully account for electroweak symmetry breaking? Does it match standard-model branching fractions to gauge bosons? In greater depth, are the absolute couplings to $W$ and $Z$ as expected in the standard model?
\item Is the Higgs field the only source of fermion masses? Are the fermion couplings proportional to fermion masses?
\item What role does the Higgs field play in generating neutrino masses?
\item Are all production rates as expected?
\item Can we establish or exclude decays to new particles? Does $H(125)$ act as a portal to hidden sectors? 
\item Can we find any sign of new strong dynamics or (partial) compositeness?
\item Can we establish the $HHH$ trilinear self-coupling?
\item How well can we test the notion that $H$ regulates Higgs--Goldstone scattering, i.e., tames the high-energy behavior of $WW$ scattering?
\item What is the order of the electroweak phase transition?
\setcounter{runningcount}{\value{enumi}}
\end{enumerate}
The last four entries call for sensitive studies at high energies---almost certainly higher than we have available at the LHC. 

\section{More new physics on the TeV scale and beyond?}
Before experiments began at the LHC, there was much informed speculation---but no guarantees---about what might be found, beyond the keys to electroweak symmetry breaking. The targets included supersymmetry and technicolor, either of which could have been a once-and-done solution to enforcing the large hierarchy between the electroweak scale and the unification scale or Planck scale. We were also encouraged by the observation that a dark-matter candidate in the form of a weakly interacting massive particle would naturally reproduce (what we take to be) the observed relic density, if the WIMP mass lay in the range of a few hundred GeV. We cannot prove that an apparently stable particle produced in the collider environment has a cosmological lifetime, but if we were to produce a candidate we could explore its properties in much greater detail than we imagine doing in direct- or indirect-detection experiments. If our reading of the evidence for dark matter is correct, we will need to assemble evidence from all the experimental approaches. No direct sign of new physics beyond the standard model has come to light, but searches must continue at the LHC and beyond~\footnote{The first hints may come from precision measurements (think of the $(g - 2)_\mu$ anomaly or the hints of lepton nonuniversality) rather than the direct observation of new phenomena.}. These considerations invite further questions.
\begin{enumerate}
\setcounter{enumi}{\value{runningcount}}
\item Do quarks and leptons show signs of compositeness? Are they made of more elementary constituents?
\item Can we find evidence of a dark matter candidate?
\item Why is empty space so nearly massless? What is the resolution to the vacuum energy problem?
\item Will ``missing energy'' events signal the existence of spacetime dimensions beyond the familiar $3+1$?
\item Can we probe dark energy in laboratory experiments?
\item Can we find clues to the origin of electroweak symmetry breaking? Is there a dynamical origin to the ``Higgs potential?''
\item Might we find indirect evidence for a new family of strongly interacting particles, such as those that are present in supersymmetric extensions of the standard model, by seeing a change in the evolution of the strong coupling ``constant,'' $1/\alphas$, at the HE-LHC or a ``100-TeV'' collider?
\item How can we constrain---or provide evidence for---light dark-matter particles or other denizens of the dark in high-energy colliders or beam-dump experiments?
\item Does the gluon have heavy partners, indicating that QCD is part of a structure richer than $\mathrm{SU(3)_c}$?
\item How can technologies developed for accelerators advance the search for axions?
\setcounter{runningcount}{\value{enumi}}
\end{enumerate}

\section{Flavor: the problem of identity}
In distinction to the issue of electroweak symmetry breaking, for which the central questions were clearly articulated for many years and we identified the 1-TeV scale as the promised land for finding answers, we do not have a clear view of how to approach the diverse character of the constituents of matter---the quarks and leptons. To be sure, we have challenged the Cabibbo--Kobayashi--Maskawa (quark-mixing matrix) paradigm and found it an extraordinarily reliable framework in the hadron sector. It is striking that, of all the parameters of the standard model (there are at least twenty-six, as listed in Table~1), no fewer than twenty pertain to flavor, and we have no idea what determines them, nor at what energy scale they are  set~\footnote{In contrast, we can see how the low-energy values of the coupling parameters $\alphas$, $\alpha_{\mathrm{em}}$, $\sin^2\theta_{\mathrm{W}}$ might be set by evolution from a unified theory at a high scale.}.  Even if we succeed in establishing that the Higgs mechanism as embodied in the electroweak theory explains \emph{how} the masses and mixing angles arise, we will not know \emph{why} they have the values we observe. That is \emph{physics beyond the standard model,} even in the case of the electron mass! What is the meta-question that underlies all these parameters~\footnote{According to the string-landscape point of view, the values might not have any deep significance.}? 
\begin{table}
\centering
{Table 1. Parameters of the Standard Model \strut}
\begin{ruledtabular}
{\begin{tabular}{@{}cl@{}}
3 & Coupling parameters, $\alphas$, $\alpha_{\mathrm{em}}$, $\sin^2\theta_{\mathrm{W}}$ \\[3pt]
2 & Parameters of the Higgs potential \\
1 & Vacuum phase (QCD) \\[3pt]
6 & Quark masses \\
3 & Quark mixing angles \\
1 & CP-violating phase\\
3 & Charged-lepton masses \\
3 & Neutrino masses \\
3 & Leptonic mixing angles \\
1 & Leptonic CP-violating phase (+ Majorana phases?)\\
\hline
$26^+$ & Arbitrary parameters\\
\end{tabular}}
\end{ruledtabular}
\label{tab:tbl1}
\end{table}

We can state one part of the problem of identity very directly: 
What makes an electron an electron and a top quark a top quark? Here are some more questions, for both theory and experiment:
\begin{enumerate}
\setcounter{enumi}{\value{runningcount}}
\item Can we find evidence of right-handed charged-current interactions? Is nature built on a fundamentally asymmetrical plan, or are the right-handed weak interactions simply too feeble for us to have observed until now, reflecting an underlying symmetry hidden by spontaneous symmetry breaking? 
\item Are there additional electroweak gauge bosons, beyond $W^\pm$ and $Z$?
\item Is charged-current universality exact? What about lepton-flavor universality?
\item Where are flavor-changing neutral currents? In the standard model, these are absent at tree level and highly suppressed by the Glashow--Iliopouolos--Maiani mechanism. They arise generically in proposals for physics beyond the standard model, and need to be controlled. And yet we have made no sightings~\footnote{The rates observed for the rare decays $K^+ \to \pi^+ \nu \bar{\nu}$ and $B_s \to \mu^+\mu^-$ are consistent with standard-model predictions.}! Why not?
\item Can we find evidence for charged-lepton flavor violation?
\item Why are there three families of quarks and leptons? (Is it so?)
\setcounter{runningcount}{\value{enumi}}
\end{enumerate}
Neutrino oscillations, discussed in the next Section, give us another take on the flavor problem. There might be more aspects to flavor. When Mendele'ev devised his periodic table, he knew nothing of the noble gases. Might there be undiscovered matter constituents that will open our eyes to some pattern?

\section{Some outstanding questions in neutrino physics}
The discovery that neutrinos oscillate among the three known species, $\nu_e, \nu_\mu, \nu_\tau$---made, incidentally, with neutrinos from natural sources---is one of the great achievements of particle physics in the past two decades. Accelerator-based experiments are playing an essential role in following up the discovery, and neutrino superbeams generated by meson decay at J-PARC and Fermilab will require proton power approaching a megawatt. The mammoth DUNE and Hyper-Kamiokande detectors as well as new short-baseline experiments are expected to begin operation in the next decade~\footnote{The tritium $\beta$-decay experiment KATRIN and nonaccelerator experiments that rely on reactors (such as JUNO) or natural sources (such as the neutrino telescopes IceCube and KM3Net) enrich the neutrino campaign.}. Among the questions we would like to answer are these:
\begin{enumerate}
\setcounter{enumi}{\value{runningcount}}
\item What is the order of levels of the mass eigenstates $\nu_1, \nu_2, \nu_3$? It is known that the $\nu_e$-rich $\nu_1$ is the lighter of the ``solar pair,'' with the more massive $\nu_2$. Does the $\nu_e$-poor $\nu_3$ lie above (``normal'' mass hierarchy) or below (``inverted hierarchy'') the others?
\item What is the absolute mass scale of neutrino masses?
\item What is the flavor composition of $\nu_3$? Is it richer in $\nu_\mu$ or $\nu_\tau$?
\item Is CP violated in neutrino oscillations? To what degree?
\item Are neutrinos Majorana particles? While this issue is primarily addressed by searches for neutrinoless double-$\beta$ decay, collider searches for same-sign lepton pairs also speak to it.
\item Do three light (left-handed) neutrinos suffice?
\item Are there light sterile neutrinos? If so, how could they arise?
\item Do neutrinos have nonstandard interactions, beyond those mediated by $W^\pm$ and $Z$?
\item How can we detect the cosmic neutrino background?
\item Are all the neutrinos stable?
\item Do neutrinos contribute appreciably to the dark matter of the Universe?
\item In what way is neutrino mass a sign of physics beyond the standard model?
\item Will neutrinos give us insight into the matter excess in the Universe (through leptogenesis)?
\setcounter{runningcount}{\value{enumi}}
\end{enumerate}
A Neutrino Factory based on a muon storage ring could provide a very strong second act for the coming generation of accelerator-based neutrino experiments. Beyond its application to oscillation experiments as an intense source with known composition, an instrument that delivered $10^{20}~\nu$ per year could be a highly valuable resource for on-campus experiments. Neutrino interactions on thin targets, polarized targets, or active targets could complement the nucleon-structure programs carried out in electron scattering at Jefferson Lab and elsewhere.

\section{An exercise for the reader}
I have now posed \arabic{runningcount} questions for experiments, and for theory that engages with experiment---by no means an exhaustive list. I offer a penultimate question for you:
\begin{enumerate}
\setcounter{enumi}{\value{runningcount}}
\item How would \emph{you} assess the scientific potential (in view of cost and schedule) of 

(a) The High-Luminosity LHC?

(b) The High-Energy LHC?

(c) A 100-TeV $pp$ Collider (FCC-hh or SppC)?

(d) A 250-GeV ILC?

(e) A circular Higgs factory (FCC-ee or CEPC)?

(f) A 380-GeV CLIC?

(g) LHeC / FCC-eh? (or an electron--ion collider optimized for nucleon studies)

(h) A muon-storage-ring neutrino factory?

(i) A multi-TeV muon collider?

(j) Another instrument of your dreams?
\setcounter{runningcount}{\value{enumi}}
\end{enumerate}
\section{Final Remarks}
The progress of accelerator science and technology has driven the development of particle physics, while the imperatives of experimental research have stimulated advances in accelerator research. I am confident that the synergy will continue. The new machines under discussion have compelling scientific motivations and appear achievable: they will require significant technological progress, but they do not invoke miracles. Together with the detectors that our experimental colleagues mount to exploit them, they are exemplars of the most amazing achievements of human beings---all the more admirable for being dedicated to the advancement of knowledge. We do not seek to build these machines out of mere habit, but because the scientific frontiers they will open are incredibly exciting. While we do our best to predict what new understanding the next accelerators will yield, the real thrill is that we don't know what we will find.

I suspect that every generation has wondered whether the next machine will be the last. Even if timely innovations in the past have pushed the boundaries of what we can do, that anxiety will always be present. The size, complexity, cost, and time scale of the accelerators we would like to attempt next amplify the concern. The long duration of projects that may not come to fruition means that the particle-physics community has a special responsibility to nurture the careers of accelerator designers and builders. That responsibility falls naturally to the great laboratories, but more university physics and engineering departments should see accelerator science as a fertile intellectual discipline, with lively connections to many other fields~\cite{AFAF}. Breakthroughs and refinements in accelerator technology may find their first---or most consequential---applications far from the frontiers that preoccupy particle physics.

The future machines that I have mentioned lie near the edge of practicality in terms of performance and resources required. It is liberating---and important---for us to look beyond projects we can credibly propose today and to dream for the far future. A good model can be found in James Bjorken's 1982 lectures on storage rings to attain $\sqrt{s}=1000\tev$~\cite{Bjorken:1982iw}! Imagine the possibilities if wake-field acceleration or some other innovation would allow us to reach gradients of many GeV---even a TeV---per meter. How would we first apply that bit of magic, and what characteristics other than gradient would be required? If we \emph{could} shrink the dimensions of multi-TeV accelerators, is there any prospect for shrinking the dimensions of detectors that depend on particle interactions with matter? What could we do with a low-emittance, high-intensity muon source? What inventions would it take to accelerate beams of particles with picosecond lifetimes? How can we imagine going far beyond current capabilities for steering beams? How could we apply high-transmissivity crystal channeling, if we could perfect it? How would optimizations change if we  were able to shape superconducting magnet coils out of biplanar graphene or an analogous material? 

This kind of talk brings us to a final question---for now:
\begin{enumerate}
\setcounter{enumi}{\value{runningcount}}
\item How are we prisoners of conventional thinking, and how can we break out?
\setcounter{runningcount}{\value{enumi}}
\end{enumerate}
The next move is yours!
\section*{Acknowledgments}
I thank Alex Chao and Weiren Chou for their kind invitation to contribute to \emph{RAST10}, and commend them for their stewardship over a decade of the journal---a remarkable resource for everyone touched by accelerator science and technology. I have benefited from insightful advice from the \emph{RAST} referees.
This manuscript has been authored by Fermi Research Alliance, LLC under Contract No. DE-AC02-07CH11359 with the U.S. Department of Energy, Office of Science, Office of High Energy Physics.

%
%
%

\bibliography{CQRAST10rt}

\begin{thebibliography}{18}%
\makeatletter
\providecommand \@ifxundefined [1]{%
 \@ifx{#1\undefined}
}%
\providecommand \@ifnum [1]{%
 \ifnum #1\expandafter \@firstoftwo
 \else \expandafter \@secondoftwo
 \fi
}%
\providecommand \@ifx [1]{%
 \ifx #1\expandafter \@firstoftwo
 \else \expandafter \@secondoftwo
 \fi
}%
\providecommand \natexlab [1]{#1}%
\providecommand \enquote  [1]{``#1''}%
\providecommand \bibnamefont  [1]{#1}%
\providecommand \bibfnamefont [1]{#1}%
\providecommand \citenamefont [1]{#1}%
\providecommand \href@noop [0]{\@secondoftwo}%
\providecommand \href [0]{\begingroup \@sanitize@url \@href}%
\providecommand \@href[1]{\@@startlink{#1}\@@href}%
\providecommand \@@href[1]{\endgroup#1\@@endlink}%
\providecommand \@sanitize@url [0]{\catcode `\\12\catcode `\$12\catcode
  `\&12\catcode `\#12\catcode `\^12\catcode `\_12\catcode `\%12\relax}%
\providecommand \@@startlink[1]{}%
\providecommand \@@endlink[0]{}%
\providecommand \url  [0]{\begingroup\@sanitize@url \@url }%
\providecommand \@url [1]{\endgroup\@href {#1}{\urlprefix }}%
\providecommand \urlprefix  [0]{URL }%
\providecommand \Eprint [0]{\href }%
\providecommand \doibase [0]{http://dx.doi.org/}%
\providecommand \selectlanguage [0]{\@gobble}%
\providecommand \bibinfo  [0]{\@secondoftwo}%
\providecommand \bibfield  [0]{\@secondoftwo}%
\providecommand \translation [1]{[#1]}%
\providecommand \BibitemOpen [0]{}%
\providecommand \bibitemStop [0]{}%
\providecommand \bibitemNoStop [0]{.\EOS\space}%
\providecommand \EOS [0]{\spacefactor3000\relax}%
\providecommand \BibitemShut  [1]{\csname bibitem#1\endcsname}%
\let\auto@bib@innerbib\@empty
\bibitem [{\citenamefont {Barber}\ \emph {et~al.}(1966)\citenamefont {Barber},
  \citenamefont {Gittelman}, \citenamefont {O'Neill},\ and\ \citenamefont
  {Richter}}]{Barber:1966kd}%
  \BibitemOpen
  \bibfield  {author} {\bibinfo {author} {\bibfnamefont {W.~C.}\ \bibnamefont
  {Barber}}, \bibinfo {author} {\bibfnamefont {Bernard~J.}\ \bibnamefont
  {Gittelman}}, \bibinfo {author} {\bibfnamefont {G.~K.}\ \bibnamefont
  {O'Neill}}, \ and\ \bibinfo {author} {\bibfnamefont {Burton}\ \bibnamefont
  {Richter}},\ }\bibfield  {title} {\enquote {\bibinfo {title} {{A Test of
  Quantum Electrodynamics by electron--electron Scattering}},}\ }\href
  {\doibase 10.1103/PhysRevLett.16.1127} {\bibfield  {journal} {\bibinfo
  {journal} {Phys. Rev. Lett.}\ }\textbf {\bibinfo {volume} {16}},\ \bibinfo
  {pages} {1127--1130} (\bibinfo {year} {1966})}\BibitemShut {NoStop}%
\bibitem [{\citenamefont {Livingston}(1969)}]{Livingston:PABH}%
  \BibitemOpen
  \bibfield  {author} {\bibinfo {author} {\bibfnamefont {Milton~Stanley}\
  \bibnamefont {Livingston}},\ }\href@noop {} {\emph {\bibinfo {title}
  {{Particle Accelerators: A Brief History}}}}\ (\bibinfo  {publisher} {Harvard
  University Press},\ \bibinfo {address} {Cambridge, Mass.},\ \bibinfo {year}
  {1969})\BibitemShut {NoStop}%
\bibitem [{\citenamefont {Chao}\ \emph {et~al.}(2013)\citenamefont {Chao},
  \citenamefont {Mess}, \citenamefont {Tigner},\ and\ \citenamefont
  {Zimmermann}}]{Handbook}%
  \BibitemOpen
  \bibfield  {author} {\bibinfo {author} {\bibfnamefont {A.~W.}\ \bibnamefont
  {Chao}}, \bibinfo {author} {\bibfnamefont {K.-H.}\ \bibnamefont {Mess}},
  \bibinfo {author} {\bibfnamefont {M.}~\bibnamefont {Tigner}}, \ and\ \bibinfo
  {author} {\bibfnamefont {F.}~\bibnamefont {Zimmermann}},\ }\href@noop {}
  {\emph {\bibinfo {title} {Handbook of Accelerator Physics and
  Engineering}}},\ \bibinfo {edition} {2nd}\ ed.\ (\bibinfo  {publisher} {World
  Scientific Publishing Company},\ \bibinfo {address} {Singapore},\ \bibinfo
  {year} {2013})\ \bibinfo {note} {\S1.6. See also the Wikipedia list of
  accelerators in particle physics,
  \url{https://en.wikipedia.org/wiki/List_of_accelerators_in_particle_physics}.
  Colliders are tabulated in M. Tanabashi et al. (Particle Data Group),
  ``Review of Particle Physics,'' Phys. Rev. D \textbf{98}, 030001 (2018), §31,
  \url{http://pdg.lbl.gov/2018/reviews/
  rpp2018-rev-hep-collider-params.pdf}.}\BibitemShut {Stop}%
\bibitem [{\citenamefont {Galison}(1997)}]{ImageLogic}%
  \BibitemOpen
  \bibfield  {author} {\bibinfo {author} {\bibfnamefont {P.~L.}\ \bibnamefont
  {Galison}},\ }\href@noop {} {\emph {\bibinfo {title} {Image and Logic: A
  Material Culture of Microphysics}}}\ (\bibinfo  {publisher} {University of
  Chicago Press},\ \bibinfo {year} {1997})\BibitemShut {NoStop}%
\bibitem [{\citenamefont {Froidevaux}\ and\ \citenamefont
  {Sphicas}(2006)}]{Froidevaux2006}%
  \BibitemOpen
  \bibfield  {author} {\bibinfo {author} {\bibfnamefont {Daniel}\ \bibnamefont
  {Froidevaux}}\ and\ \bibinfo {author} {\bibfnamefont {Paris}\ \bibnamefont
  {Sphicas}},\ }\bibfield  {title} {\enquote {\bibinfo {title} {General-purpose
  detectors for the large hadron collider},}\ }\href {\doibase
  10.1146/annurev.nucl.54.070103.181209} {\bibfield  {journal} {\bibinfo
  {journal} {Annual Review of Nuclear and Particle Science}\ }\textbf {\bibinfo
  {volume} {56}},\ \bibinfo {pages} {375--440} (\bibinfo {year}
  {2006})}\BibitemShut {NoStop}%
\bibitem [{\citenamefont {Campana}\ \emph {et~al.}(2016)\citenamefont
  {Campana}, \citenamefont {Klute},\ and\ \citenamefont {Wells}}]{Campana2016}%
  \BibitemOpen
  \bibfield  {author} {\bibinfo {author} {\bibfnamefont {P.}~\bibnamefont
  {Campana}}, \bibinfo {author} {\bibfnamefont {M.}~\bibnamefont {Klute}}, \
  and\ \bibinfo {author} {\bibfnamefont {P.~S.}\ \bibnamefont {Wells}},\
  }\bibfield  {title} {\enquote {\bibinfo {title} {Physics goals and
  experimental challenges of the proton{\textendash}proton high-luminosity
  operation of the {LHC}},}\ }\href {\doibase
  10.1146/annurev-nucl-102115-044812} {\bibfield  {journal} {\bibinfo
  {journal} {Annual Review of Nuclear and Particle Science}\ }\textbf {\bibinfo
  {volume} {66}},\ \bibinfo {pages} {273--295} (\bibinfo {year} {2016})},\
  \bibinfo {note} {{Also see the work carried out by the International Linear
  Collider Detector R\&D Groups,
  \url{http://www.linearcollider.org/P-D/Detector-R-D-groups}}}\BibitemShut
  {NoStop}%
\bibitem [{\citenamefont {Bird}\ \emph {et~al.}(2014)\citenamefont {Bird} \emph
  {et~al.}}]{Bird:1695401}%
  \BibitemOpen
  \bibfield  {author} {\bibinfo {author} {\bibfnamefont {I.}~\bibnamefont
  {Bird}} \emph {et~al.},\ }\href@noop {} {\enquote {\bibinfo {title} {{Update
  of the Computing Models of the WLCG and the LHC Experiments}},}\ } (\bibinfo
  {year} {2014}),\ \bibinfo {note}
  {\url{https://cds.cern.ch/record/1695401}}\BibitemShut {NoStop}%
\bibitem [{\citenamefont {Bauerdick}\ \emph {et~al.}(2018)\citenamefont
  {Bauerdick} \emph {et~al.}}]{Bauerdick:2018qjx}%
  \BibitemOpen
  \bibfield  {author} {\bibinfo {author} {\bibfnamefont {Lothar}\ \bibnamefont
  {Bauerdick}} \emph {et~al.},\ }\href@noop {} {\enquote {\bibinfo {title}
  {{HEP Software Foundation Community White Paper Working Group - Data Analysis
  and Interpretation}},}\ } (\bibinfo {year} {2018}),\ \Eprint
  {http://arxiv.org/abs/1804.03983} {arXiv:1804.03983 [physics.comp-ph]}
  \BibitemShut {NoStop}%
\bibitem [{\citenamefont {Cole}\ \emph {et~al.}(1968)\citenamefont {Cole},
  \citenamefont {Goldwasser},\ and\ \citenamefont {Wilson}}]{Cole:1968wx}%
  \BibitemOpen
  \bibfield  {author} {\bibinfo {author} {\bibfnamefont {Frank~T.}\
  \bibnamefont {Cole}}, \bibinfo {author} {\bibfnamefont {Edwin~L.}\
  \bibnamefont {Goldwasser}}, \ and\ \bibinfo {author} {\bibfnamefont
  {Robert~Rathbun}\ \bibnamefont {Wilson}},\ }\href@noop {} {\emph {\bibinfo
  {title} {{National Accelerator Laboratory Design Report}}}},\ \bibinfo {type}
  {Tech. Rep.}\ (\bibinfo  {institution} {{Fermilab}},\ \bibinfo {year}
  {1968})\ \bibinfo {note}
  {\href{http://inspirehep.net/record/53213/files/fermilab-design-1968-01.pdf}{FERMILAB-DESIGN-1968-01},
  \S2.2}\BibitemShut {NoStop}%
\bibitem [{\citenamefont {Wilson}(1979)}]{Wilson:1979zg}%
  \BibitemOpen
  \bibfield  {author} {\bibinfo {author} {\bibfnamefont {R.~R.}\ \bibnamefont
  {Wilson}},\ }\bibfield  {title} {\enquote {\bibinfo {title} {{Fantasies of
  Future Fermilab Facilities}},}\ }\href {\doibase 10.1103/RevModPhys.51.259}
  {\bibfield  {journal} {\bibinfo  {journal} {Rev. Mod. Phys.}\ }\textbf
  {\bibinfo {volume} {51}},\ \bibinfo {pages} {259--273} (\bibinfo {year}
  {1979})}\BibitemShut {NoStop}%
\bibitem [{\citenamefont {Quigg}(2009)}]{Quigg:2009vq}%
  \BibitemOpen
  \bibfield  {author} {\bibinfo {author} {\bibfnamefont {Chris}\ \bibnamefont
  {Quigg}},\ }\bibfield  {title} {\enquote {\bibinfo {title} {{Unanswered
  Questions in the Electroweak Theory}},}\ }\href {\doibase
  10.1146/annurev.nucl.010909.083126} {\bibfield  {journal} {\bibinfo
  {journal} {Ann. Rev. Nucl. Part. Sci.}\ }\textbf {\bibinfo {volume} {59}},\
  \bibinfo {pages} {505--555} (\bibinfo {year} {2009})},\ \bibinfo {note}
  {{arXiv:0905.3187 [hep-ph]; and ``Electroweak Symmetry Breaking in Historical
  Perspective,'' \textbf{65}, 25--42 (2015), arXiv:1503.01756
  [hep-ph]}}\BibitemShut {NoStop}%
\bibitem [{Note1()}]{Note1}%
  \BibitemOpen
  \bibinfo {note} {The first hints may come from precision measurements (think
  of the $(g - 2)_\mu $ anomaly or the hints of lepton nonuniversality) rather
  than the direct observation of new phenomena.}\BibitemShut {Stop}%
\bibitem [{Note2()}]{Note2}%
  \BibitemOpen
  \bibinfo {note} {In contrast, we can see how the low-energy values of the
  coupling parameters $\protect \ensuremath {\alpha _{\protect \mathrm {s}}}$,
  $\alpha _{\protect \mathrm {em}}$, $\protect \qopname \relax o{sin}^2\theta
  _{\protect \mathrm {W}}$ might be set by evolution from a unified theory at a
  high scale.}\BibitemShut {Stop}%
\bibitem [{Note3()}]{Note3}%
  \BibitemOpen
  \bibinfo {note} {According to the string-landscape point of view, the values
  might not have any deep significance.}\BibitemShut {Stop}%
\bibitem [{Note4()}]{Note4}%
  \BibitemOpen
  \bibinfo {note} {The rates observed for the rare decays $K^+ \to \pi ^+ \nu
  \protect \mathaccentV {bar}016{\nu }$ and $B_s \to \mu ^+\mu ^-$ are
  consistent with standard-model predictions.}\BibitemShut {Stop}%
\bibitem [{Note5()}]{Note5}%
  \BibitemOpen
  \bibinfo {note} {The tritium $\beta $-decay experiment KATRIN and
  nonaccelerator experiments that rely on reactors (such as JUNO) or natural
  sources (such as the neutrino telescopes IceCube and KM3Net) enrich the
  neutrino campaign.}\BibitemShut {Stop}%
\bibitem [{AFA(2009)}]{AFAF}%
  \BibitemOpen
  \href@noop {} {\enquote {\bibinfo {title} {{Accelerators for America's
  Future}},}\ } (\bibinfo {year} {2009}),\ \bibinfo {note}
  {{\href{https://science.energy.gov/~/media/hep/pdf/accelerator-rd-stewardship/Report.pdf}{U.S.
  Department of Energy Report,
  \url{http://science.energy.gov/~/media/hep/pdf/accelerator-rd-stewardship/Report.pdf}.
  See also \url{http://www.acceleratorsamerica.org}}}}\BibitemShut {NoStop}%
\bibitem [{\citenamefont {Bjorken}(1983)}]{Bjorken:1982iw}%
  \BibitemOpen
  \bibfield  {author} {\bibinfo {author} {\bibfnamefont {J.~D.}\ \bibnamefont
  {Bjorken}},\ }\enquote {\bibinfo {title} {{A Thousand TeV in the
  Center-of-Mass: Introduction to High-Energy Storage Rings}},}\ in\ \href
  {\doibase 10.1007/978-1-4613-3745-4_6} {\emph {\bibinfo {booktitle}
  {{Techniques and Concepts of High-Energy Physics: Proceedings, 2nd NATO
  Advanced Study Institute, Lake George, New York, July 1-12, 1982}}}},\
  Vol.~\bibinfo {volume} {99},\ \bibinfo {editor} {edited by\ \bibinfo {editor}
  {\bibfnamefont {T.}~\bibnamefont {Ferbel}}}\ (\bibinfo {year} {1983})\ pp.\
  \bibinfo {pages} {233--300},\ \bibinfo {note} {FERMILAB-CONF-82-055-THY,
  \url{https://link.springer.com/content/pdf/10.1007/978-1-4613-3745-4.pdf}}\BibitemShut
  {NoStop}%
\end{thebibliography}%

\end{document}